\documentclass[%
 reprint,
 amsmath,amssymb,
 aps,
]{revtex4-2}
\usepackage{graphicx}
\usepackage{dcolumn}
\usepackage{bm}

\usepackage[normalem]{ulem}
\usepackage{color,xcolor}
\usepackage{graphicx}
\usepackage[colorlinks]{hyperref}

\begin{document}

\preprint{APS/123-QED}

\title{Intersubband electric dipole spin resonance in transition metal dichalcogenide heterobilayers
}

\author{K.K. Grigoryan}
\affiliation{L. D. Landau Institute for Theoretical Physics, 142432 Chernogolovka, Russia}

\author{M.M. Glazov}
\affiliation{Ioffe Institute, 194021 St. Petersburg, Russia }%

\date{\today}

\begin{abstract}

The theory of inter-spin-subband electric dipole spin resonance in transition metal dichalcogenide heterobilayers is proposed. Our  symmetry analysis demonstrates that, in contrast to monolayers, the reduced symmetry of heterobilayers enables coupling between conduction band spin subbands by an electric field. We establish the optical selection rules for all six high-symmetry stacking configurations. The microscopic mechanism of the effect is identified as the spin-orbit coupling induced mixing of Bloch states from different conduction bands, which generates a non-zero momentum matrix element between the spin-split states. It also leads to the linear-in-wavevector spin-dependent terms in the effective Hamiltonian, i.e.,  the Rashba effect. Our estimates show that the rate of electric-dipole spin-flip transitions exceeds by far that of the magnetic-dipole transitions in transition metal dichalcogenide heterobilayers.

\end{abstract}

\maketitle


\section{\label{sec:intro}Introduction}

Transition metal dichalcogenides (TMDCs)  monolayers and heterostructures are of significant interest for modern fundamental condensed matter physics, nanotechnology, and quantum technologies~\cite{Kolobov2016book}. Their direct bandgap in the visible to near-infrared range makes them ideal for studying excitonic effects~\cite{RevModPhys.90.021001,Durnev_2018,Glazov:2024aa}, light-matter interaction~\cite{Schneider:2018aa,Kipp:2025aa}, and opens pathways for optoelectronic applications~\cite{Thakar:2020aa}. One of the key features of TMDCs is the presence of two time reversal symmetry related valleys ($K^+$ and $K^-$) in the Brillouin zone~\cite{Xiao:2012cr,2053-1583-2-2-022001}. The strong spin-subband splitting due to spin-orbit coupling and specific optical selection rules allows for charge carrier and exciton spin and valley orientation using circularly polarized light, and valley entanglement by linearly polarized light~\cite{Mak:2012qf,Kioseoglou,PhysRevLett.112.047401,PhysRevB.92.075409,Glazov_2021,66rj-jbqw}.

Semiconducting TMDCs form the building blocks for van der Waals heterostructures~\cite{Geim:2013aa}. They provide a versatile platform for heterostructure engineering: combining materials with different bandgaps allows generally type-II band alignment~\cite{10.1063/1.4774090}, resulting in spatially separated electrons and holes, symmetry tuning, and integration with other two-dimensional materials such as graphene and hexagonal boron nitride to design structures with tailored properties~\cite{Geim:2013aa,Castellanos-Gomez:2022aa,PhysRevResearch.3.043217}. It also gives rise to fascinating manybody physics related to the formation of moir\'e patterns and manybody correlated states~\cite{Zhange1601459,Xu:2020aa,Li:2021uk,Baek:2021aa,Ciorciaro:2023aa,Tan:2023aa}.

A novel degree of freedom related to the stacking of individual layers in heterostructures makes it possible to control the symmetries and optical transition rules in heterobilayers~\cite{Yu:2018aa,Forg:2019aa}. It allows not only control of spin- and valley-dependent optical properties, but also -- combined with magnetooptical spectroscopy -- identification of the layer stacking in heterostructures~\cite{Zhao:2023aa}.

On the other hand, much less is known about the spin-dependent intersubband transitions, which are crucial for electron spin and valley physics. For instance, chiral-phonon enabled electron spin relaxation~\cite{Song:2013uq,PhysRevB.110.195403} is a key to the formation of spin-dark excitons, which usually dominate the optical spectra of tungsten-based TMDC~\cite{Wang:2017b}. More importantly, the conduction band spin-orbit splitting in TMDC monolayers and bilayers is about $10\ldots 20$~meV~\cite{PhysRevB.84.153402,PhysRevB.88.245436,PhysRevB.93.121107} corresponding to the terahertz (THz) spectral range. The THz spectroscopy serves as a versatile tool to address electronic transitions in semiconductors, including classical systems and emerging TMDCs \cite{ganichev_book,leinss2008terahertz, Poellmann:2015aa,Sie:2017aa,Yong:2019aa,Langer:2018aa,Venanzi2021,Venanzi:2024aa,39cj-24hk}. The natural question is whether the inter-spin-subband transitions can be probed by the THz radiation and, if so, what the microscopic mechanism of the effect is?

In this paper, we develop the theory of intersubband transitions in TMDC-based monolayers and heterobilayers. We perform a symmetry analysis of the effect for the high-symmetry stackings in moir\'e-free heterobilayers. We demonstrate that the spin-subband transitions in free standing monolayers are allowed only in the magnetic-dipole approximation. In contrast, heterobilayers allow both electric and magnetic-dipole transitions between spin-orbit split subbands. This enables electric dipole spin resonance (EDSR) -- an effect originally predicted by E.I. Rashba for bulk semiconductors and low dimensional structures \cite{rashbasheka, rashba64, Rashba03}, see Refs.~\cite{RASHBA:1991aa, dyakonov_book}. 

Shortly after prediction~\cite{rashbasheka}, the EDSR has been observed in bulk semiconductors~\cite{mccombeCombinedResonanceElectron1967,mccombeInfraredStudiesCombined1969,mccombeElectricDipoleExcitedElectronSpin1971}. It was quickly realized that the electric dipole mechanism dominates the spin resonance and enables combined, spin and cyclotron, resonance in wide range of systems~\cite{RASHBA:1991aa}. In addition to classical semiconductors and semiconductor quantum well and quantum dot structures, the EDSR is observed in SiGe-based nanosystems where initially it was assumed that the spin-orbit interaction is suppressed~\cite{wilamowskiEvidenceEvaluationBychkovRashba2002,wilamowskiGFactorTuningManipulation2007,zinovievaSpinResonanceElectrons2008}. Rise of graphene and other two-dimensional materials has, naturally, attracted attention to EDSR in these systems resulting in a number of theoretical predictions~\cite{brooksElectricDipoleSpin2020,kumarZerofieldSpinResonance2021,denisovTerahertzSpinlightCoupling2024}. While the spin resonance has been observed in graphene-based nanosystems~\cite{fujitaDirectObservationElectrically2016,morissetteDiracRevivalsDrive2023}, its mechanisms are insufficiently studied so far and, to the best of our knowledge, the EDSR in TMDC monolayers has not yet been observed.

The present work presents the theory of EDSR in TMDC heterobilayers. The spin-flip intersubband transitions can be studied in $n$-type heterobilayers where resident electrons are present as a result of doping under THz irradiation. They can be detected also in undoped structures in the presence of optical excitation that generates excitons: In such setting, an additional THz irradiation will cause transitions between the different types of excitons where an electron is either in the bottom or top spin subband, so called triplet and singlet excitons.

The heterobilayer studied in this work consists of two TMDC monolayers, $\mathrm{WSe_2}$ and $\mathrm{MoSe_2}$, forming a widespread type-II heterostructure with spatially separated electrons and holes. It is the prototypical system that can also be fabricated in the moir\'e free form~\cite{Barr2022,Zhao:2023aa,PhysRevLett.132.016202,PhysRevX.14.031025}.  While our calculations focus on this specific system, the results are directly applicable to other type-II heterostructures. Our primary objectives are, therefore, to: (i) establish on the symmetry basis the selection rules for spin-flip inter-subband transitions (Sec.~\ref{sec:symmetry}) and (ii) provide the model of the microscopic mechanisms governing these transitions in heterobilayers (Sec.~\ref{sec:micro}).

\section{Symmetry analysis}\label{sec:symmetry}

\subsection{Monolayer}

\begin{figure}[b]
\includegraphics[width=\linewidth]{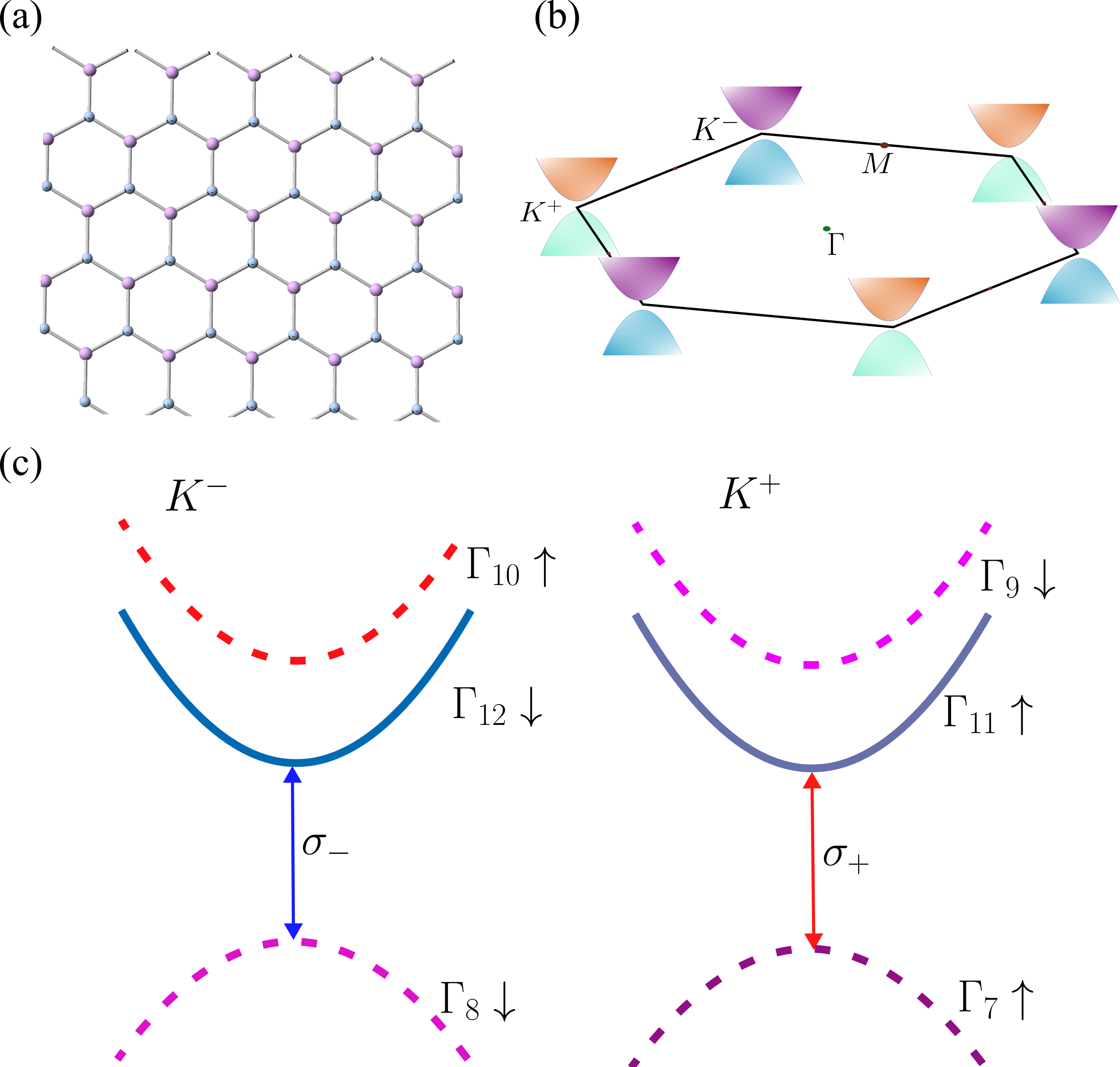}
\caption{(a) Top view of the monolayer TMDC lattice structure. Purple circles represent metal atoms, while blue circles denote the in-plane projections of chalcogen atom pairs. (b) Brillouin zone with high-symmetry points labeled. (c) Band structure close to $K^+$ and $K^-$ points, showing the highest valence band and two spin-split conduction subbands. }
\label{zone}
\end{figure}

The crystalline lattice, Brillouin zone, and electron dispersion in the vicinity of the $K^+$ and $K^-$ points of the TMDC monolayer are sketched in Fig.~\ref{zone}. The $D_{3h}$ point symmetry group of the monolayer is reduced to the $C_{3h}$ wavevector group of $K^+$ and $K^-$ points of the Brillouin zone.

The analysis~\cite{Kormanyos:2013dq,2053-1583-2-2-022001,PSSB:PSSB201552211} shows that the valence band orbital Bloch functions are invariant under point-group transformations; that is, they transform according to the invariant representation of the $A'$ in notations of Ref.~\cite{Kormanyos:2013dq} or $\Gamma_1$ in notations of Ref.~\cite{koster63}. Note that for the monolayer, the center of the point-group transformations can be placed on different lattice sites; here, we select it to be the center of the hexagon. The conduction band orbital states  in the valleys $K^+$ and $K^-$ transform according to $E'_1$ ($\Gamma_2$), i.e., as $x + \mathrm iy$ and $E'_2$ ($\Gamma_3$), as $x- \mathrm iy$, respectively. This symmetry analysis demonstrates the well-known chiral selection rules for interband optical transitions.

The inclusion of the electron spin and spin-orbit interaction results in the splitting of otherwise two-fold degenerate conduction and valence band states across the Brillouin zone and, in particular, at the $K^\pm$ points where the band extrema are realized. The valence band subbands transform according to the $\Gamma_7$ and $\Gamma_8$ irreducible representations of the $C_{3h}$ point group. The valence band splitting on the order of several hundred meV can be controlled by the composition of the system~\cite{Wang:2015aa} and manifests itself as a dominant contribution to the splitting of the so-called $A$- and $B$-transitions. Note that $\Gamma_7\times \Gamma_8^* = \Gamma_5$ and $\Gamma_7^*\times \Gamma_8=\Gamma_6$ (the bottom valence subbands are not shown in Fig.~\ref{zone}(c) because they are irrelevant for the followng) the transitions between the valence subbands are possible only in the magnetic-dipole approximation. This is because the in-plane components of a pseudovector transform according to the $\Gamma_5+\Gamma_6$ while the components of the vector transform according to $\Gamma_2+\Gamma_3$~\cite{koster63}. Due to the large energy distance, the transitions between the valence band subbands are beyond the scope of the present work.

Depending on the valley, the conduction band states transform according to the $\Gamma_{11}$ and $\Gamma_9$ (spin-up and spin-down states in the $K^+$ valley) or $\Gamma_{12}$ and $\Gamma_{10}$ (spin-down and spin-up states in the $K^-$ valley). The products $\Gamma_{11}\times \Gamma_{9}^*=\Gamma_5$ and $\Gamma_{12}\times \Gamma_{10}^*=\Gamma_6$, i.e., the THz transitions between the conduction subbands, are allowed in the magnetic-dipole approximation only. Hence, despite the lack of space inversion and strong spin-orbit coupling, the electron dipole spin resonance in the monolayers is forbidden. 

Note that an external out-of-the-plane electric field or the presence of the substrate reduces the symmetry further and allows for electric-dipole coupling between the conduction subbands in the monolayers due to the field-induced structure inversion asymmetry (Rashba effect)~\cite{PhysRevX.4.011034}. The ana\-ly\-sis of the structure inversion asymmetry in monolayers is beyond the scope of the paper.

\subsection{Heterobilayer}

\begin{figure}[h!]
\includegraphics[width=0.9\linewidth]{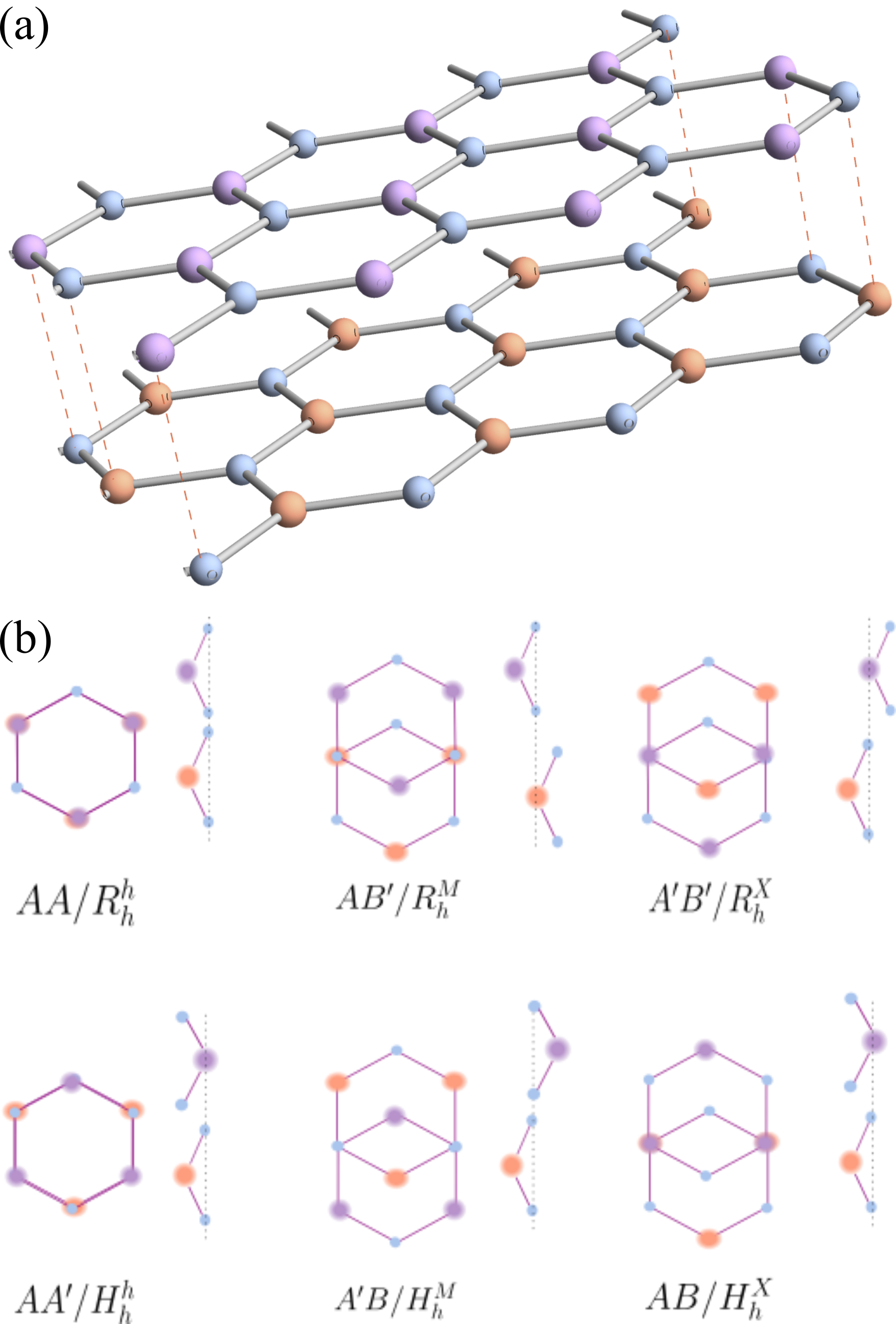}
\caption{\label{registry} 
(a) Illustration of heterobilayer in $AA/R_h^h$ stacking. Orange and magenta balls show transition metal atoms. For clarity two chalcogen atoms are represented by one blue ball. (b) Various possible high-symmetry stacking configurations of two monolayers. In \textit{H}-type stackings, the layers are rotated relative to each other, while \textit{R}-type stackings maintain unrotated layer alignment. 
}
\end{figure}

The presence of two layers in heterostructures reduces the symmetry and enables various twist configurations between the layers. Hereafter we neglect small mismatch of the lattice constants and assume an ideal alignement. There exist six high-symmetry stacking arrangements (registries), categorized into two fundamental types:
\emph{H}-type ($AA'$ or $H_h^h$, $A'B$ or $H_h^M$, $AB$ or $H_h^X$) layers rotated by $180^\circ$ relative to each other and \emph{R}-type ($AA$ or $R_h^h$, $AB'$ or $R_h^M$, $A'B'$ or $R_h^X$) layers without relative rotation~\cite{Yu:2018aa,Forg:2019aa,Zhao:2023aa,PhysRevLett.132.016202}. These distinct stacking configurations are illustrated in Fig.~\ref{registry}.

\begin{figure}[h!]
\includegraphics[width=0.9\columnwidth]{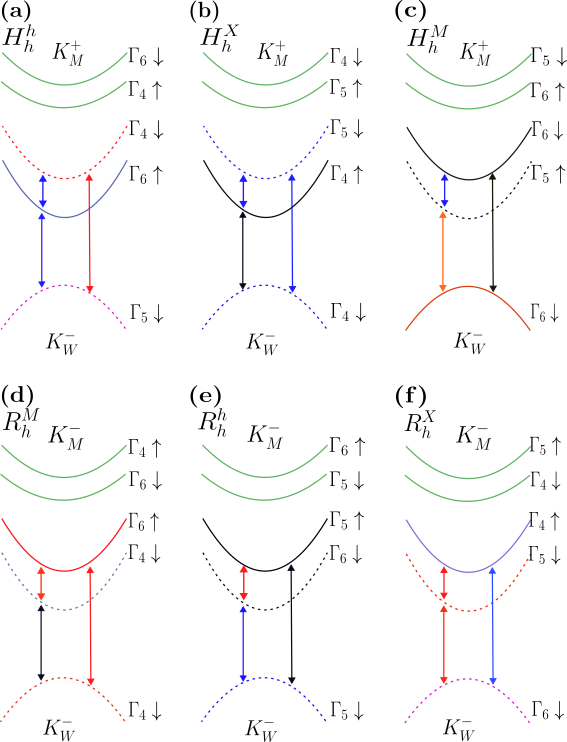}\caption{\label{dpm} 
Band structure in a single valley for all six high-symmetry stackings, showing the highest valence subband ($v$), two spin-split conduction subbands ($c$), and symmetry-equivalent spin-split subbands from the first ($c+1$) and second ($c+2$) excited conduction bands (additional spin subbands omitted for clarity) in the same layer. Energy splittings and curvatures of the bands are not to scale. Transition selection rules are indicated by colored arrows: red for $\sigma^+$ polarization (in-plane electric field), blue for $\sigma^-$ polarization (in-plane), and black for $z$ polarization (out-of-plane electric field). }
\end{figure}

Figure~\ref{dpm} schematically illustrates the band structures for all six stacking configurations, showing states in a single valley. Typically in the MoX$_2$/WY$_2$ heterostructure, where X,Y=S or Se the lowest conduction band states are in the Mo-based monolayer while the topmost valence band states are in the W-based monolayer. We note that in \emph{H}-type stackings, the $K^+$ in the MoX$_2$ layer corresponds to the $K^-$ in the W$Y_2$ layer and vice versa, whereas \emph{R}-type stackings maintain identical valley alignment across both layers. It is due to the $60^\circ$ (or, equivalently, $180^\circ$) rotation of the layers in the \emph{H}-stacking. In all cases, we consider just one valley that corresponds to the $K^-$ valley in the W-based layer. The valence band ($v$) is represented by only its highest-energy subband since, as mentioned above, the lower spin subband is irrelevant due to its several-hundred-meV separation via spin-orbit coupling. The plotted conduction band states include the two lowest-energy spin-subbands ($c$), along with two symmetry-equivalent states from the first excited ($c+1$) and the next higher ($c+2$) conduction bands stemming from the same layer bandstructure, both essential for the microscopic description of the effect, see Sec.~\ref{sec:micro} below. Under the reflection symmetry of an isolated monolayer, the $v$, $c$, and $c+2$ orbital states are invariant (their Bloch functions are even functions of the out-of-plane coordinate), while the $c+1$ orbital states are odd~\cite{2053-1583-2-2-022001,Durnev_2018}.

The symmetry of the heterobilayer is described by the $C_{3v}$ point symmetry group. Indeed, the horizontal reflection plane vanishes since the layers in the heterostructure are different; thus, both the in-plane $C_2$ axes and the mirror-reflection $S_3$ axes that were present in the monolayer's $D_{3h}$ symmetry group are absent. The remaining elements are the vertical $C_3$ axes and $\sigma_v$ reflection planes containing the $C_3$ axis, in addition to the identity operation. Consequently, the symmetry of the valley is reduced to $C_3$, where only the identity and three-fold rotation remain. The relevant spinor representations of the conduction and valence  band states depend on the stacking, which determines the position of the point-transformation center (to be common for both layers), and can be found using the method developed in Ref.~\cite{Forg:2019aa}. The results are summarized in Tab.~\ref{tab:irreps}.

\begin{table}
    \centering
    \caption{
 Symmetry classification of spin-split subbands in the lowest three conduction bands and uppermost valence band. Irreducible representations are labeled in the notations of Ref.~\cite{koster63} for the $C_3$ wavevector group.
}
    \begin{tabular}{c|c|c|c|c} \hline \hline
         registry &  $c_{\downarrow}/c_{\uparrow}$&  $\left(c+1\right)_{\downarrow}/\left(c+1\right)_{\uparrow}$&  $\left(c+2\right)_{\downarrow}/\left(c+2\right)_{\uparrow}$& $v_{\downarrow}/v_{\uparrow}$\\ \hline \hline 
         $H_h^X$~or~$AB$&  $\Gamma_{5}/\Gamma_{4}$&  $\Gamma_{6}/\Gamma_{5}$&  $\Gamma_{4}/\Gamma_{6}$& $\Gamma_{4}/\Gamma_{6}$\\ \hline 
         $H_h^h$~or~$AA^{\prime}$&  $\Gamma_{4}/\Gamma_{6}$&  $\Gamma_{5}/\Gamma_{4}$&  $\Gamma_{6}/\Gamma_{5}$& $\Gamma_{5}/\Gamma_{4}$\\ \hline 
         $H_h^M$~or~$A^{\prime}B$& $ \Gamma_{6}/\Gamma_{5}$&  $\Gamma_{4}/\Gamma_{6}$&  $\Gamma_{5}/\Gamma_{4}$& $\Gamma_{6}/\Gamma_{5}$\\ \hline 
        $R_h^h$~or~ $AA$& $ \Gamma_{6}/\Gamma_{5}$&  $\Gamma_{5}/\Gamma_{4}$&  $\Gamma_{4}/\Gamma_{6}$& $\Gamma_{5}/\Gamma_{4}$\\ \hline 
         $R_h^M$~or~$AB^{\prime}$&$\Gamma_{4}/\Gamma_{6}$&  $\Gamma_{6}/\Gamma_{5}$&  $\Gamma_{5}/\Gamma_{4}$& $\Gamma_{4}/\Gamma_{6}$\\ \hline 
         $R_h^X$~or~$A^{\prime}B^{\prime}$&  $\Gamma_{5}/\Gamma_{4}$&  $\Gamma_{4}/\Gamma_{6}$&  $\Gamma_{6}/\Gamma_{5}$& $\Gamma_{6}/\Gamma_{4}$\\ \hline
    \end{tabular}
\label{tab:irreps}
\end{table}

The selection rules for transitions between relevant bands and subbands can be determined from the irreducible representations of the Bloch functions at the $K^+$ ($K^-$) points. As established in~\cite{koster63}, within the $C_3$ wavevector group, the circular polarization components transform as follows: $\sigma^+$ polarized fields ($E_x + \mathrm{i}E_y$ or $B_x + \mathrm{i}B_y$) belong to the $\Gamma_2$ representation, while $\sigma^-$ polarized fields ($E_x - \mathrm{i}E_y$ or $B_x - \mathrm{i}B_y$) correspond to $\Gamma_3$. Here, the $x,y$ axes lie in the heterobilayer plane, with $z$ being the surface normal. Crucially, the $C_3$ group does not distinguish in-plane components of the vector and pseudovector. Since both electric and magnetic field components transform according to equivalent irreducible representations of $C_3$, transitions allowed in the magnetic-dipole approximation are also allowed in the electric-dipole approximation. This conclusion also holds for the transitions between the valence band subbands: in all studied stackings these transitions are allowed in the electron-dipole approximation.

The selection rules are indicated by color-coded arrows in Fig.~\ref{dpm}. While the interband transition selection rules agree with the literature~\cite{Yu:2018aa,Forg:2019aa}, the key new result is the selection rules for transitions between spin-split conduction subbands. Remarkably, for all high-symmetry stackings studied here, the inter-subband transitions within the conduction band ($c$) are allowed in the electric-dipole approximation, with $\sigma^+$ or $\sigma^-$ polarization depending on the stacking type.

For completeness, we present a simplified analysis of the selection rules based on basic symmetry considerations. The $C_3$ point group does not distinguish between true vectors (electric field) and pseudovectors (magnetic field). Consequently, any spin-flip transition ($\uparrow\to\downarrow$) allowed in the magnetic-dipole approximation for a given polarization (e.g., $\sigma^-$) will also be permitted in the electric-dipole approximation with the same circular polarization. Furthermore, the $C_3$ symmetry imposes angular momentum conservation modulo 3. This allows the derivation of intersubband selection rules from interband transitions through the following angular momentum addition rules: $\sigma^+ + \sigma^+ = \sigma^-$, $\sigma^- + \sigma^- = \sigma^+$, $\sigma^\pm + z = \sigma^\pm$, where $z$ denotes out-of-plane polarized transitions.

\section{Microscopic Model}\label{sec:micro}

\subsection{Spin-orbit coupling and electric dipole spin resonance}

Our symmetry analysis demonstrates the possibility of intersubband EDSR in TMDC heterobilayers. We now present an illustrative microscopic model of this effect, focusing on the $AA'$ ($H_h^h$) stacking configuration as a representative case; other stackings follow analogously.

The studied  MoX$_2$/WY$_2$ heterobilayer exhibits type-II band alignment: the conduction band minimum resides in the molybdenum-based layer, while the valence band maximum localizes in the tungsten-based layer. In the absence of interlayer coupling and spin-orbit mixing, the conduction band states $\Gamma_6$ and $\Gamma_4$ at the $K^+$ point ($K^+_M$ valley) can be expressed as:
\begin{equation}
    \label{cb:simple}
|\Gamma_6\rangle = -\frac{\mathcal{X} + \mathrm{i}\mathcal{Y}}{\sqrt{2}}\uparrow, \quad |\Gamma_4\rangle = -\frac{\mathcal{X} + \mathrm{i}\mathcal{Y}}{\sqrt{2}}\downarrow,
\end{equation}
where $\mathcal{X}, \mathcal{Y}, \mathcal{Z}$ denotes orbital Bloch functions transforming as  respective coordinates $x$, $y$, and $z$, and $\uparrow$, $\downarrow$ represent spin basis states. The negative sign ensures consistency with the canonical basis. The orbital part $\propto x+\mathrm i y$ is in agreement with the $K^+$ valley conduction band symmetry in the monolayer. Equation~\eqref{cb:simple} reveals that for an isolated monolayer, the inter-spin-subband transitions are only allowed in the magnetic-dipole approximation via the Zeeman-like coupling
\begin{equation}
    \label{Zeeman}
    \mathcal{H}_B = \frac{1}{2}g\mu_B(\bm{\sigma} \cdot \bm{B}),
\end{equation}
where $\bm{B} = (B_x,B_y)$ is the in-plane magnetic field component, $\mu_B$ is the Bohr magneton, and $g \approx 2$ is the in-plane electron $g$-factor~\cite{Robert:2020tw}.

The allowance for the interband matrix elements of the spin-orbit interaction results in the mixing of Bloch bands with different symmetry~\cite{pikusmarushaktitkov_eng,Wang:2017b,PhysRevB.93.121107}. The analysis shows that in heterobilayers, spin-orbit coupling induces state mixing as follows:
\begin{subequations}
    \label{cb:mixed}
    \begin{align}
    |\Gamma_4\rangle &= -\beta\frac{\mathcal{X} + \mathrm{i}\mathcal{Y}}{\sqrt{2}}\downarrow + \underbrace{\alpha\mathcal{Z}\uparrow}_{c+1}, \\
    |\Gamma_6\rangle &= -\beta'\frac{\mathcal{X} + \mathrm{i}\mathcal{Y}}{\sqrt{2}}\uparrow + \underbrace{\alpha'\frac{\mathcal{X}' - \mathrm{i}\mathcal{Y}'}{\sqrt{2}}\downarrow}_{c+2}.
    \end{align}
\end{subequations}
In Eqs.~\eqref{cb:mixed}, $\mathcal X'$, and $\mathcal Y'$ are the Bloch amplitudes of the $c+2$ band, while $\mathcal Z$ is the Bloch amplitude of the $c+1$ band, $\alpha$, $\alpha'$, and $\beta$, $\beta'$ are the coefficients related by the normalization constraints:
\begin{equation}
    \label{norm}
    |\alpha|^2 + |\beta|^2 = 1, \quad |\alpha'|^2 + |\beta'|^2 = 1.
\end{equation}
Equations~\eqref{cb:mixed} are derived for the $K^+$ valley, the states in the $K^-$ valley can be obtained by time reversal. In the absence of the band mixing $\beta=1$, $\alpha=\alpha'=0$. Coefficient $\alpha\ne 0$ already in the monolayers. Density functional theory calculations~\cite{PhysRevB.93.121107,PhysRevB.110.195403} in agreement with the tight-binding approach~\cite{PhysRevB.88.245436} yield $|\alpha|^2 \approx 0.02$. The microscopic estimate of the value of $\alpha'$ (which is non-zero in heterobilayers only) is unknown, but it can be expected to have the same order of magnitude as $\alpha$. Microscopically, it can be related to the state mixing caused by the interlayer coupling, and to the spin-orbit mixing with $c+2$ conduction via intermediate state with $2^{-1/2}(\mathcal X - \mathrm i \mathcal Y)\mathcal Z$ Bloch amplitude~\cite{PhysRevB.110.195403}. As we see below, the oscillator strength of the intersubband transition scales as $|\alpha|^2$ ($|\alpha'|^2$).

The mixed states~\eqref{cb:mixed} enable electric-dipole transitions. The electron momentum operator matrix element:
\begin{equation}
    \label{momentum}
    p_{\downarrow\uparrow} = \langle \Gamma_4|\bm{e}_-^* \cdot \hat{\bm{p}}|\Gamma_6\rangle 
    = -\alpha^*\beta'\gamma_{xz}
    - \alpha'\beta^*\gamma_6,
\end{equation}
with $\bm e_- = (\hat{\bm x} - \mathrm i \hat{\bm y})/\sqrt{2}$ being the unit vector of circular polarization, contains two non-vanishing contributions obtained from expanding $|\Gamma_4\rangle$ and $|\Gamma_6\rangle$ states into corresponding Bloch wavefunction  according to Eqs.~\eqref{cb:mixed}: (i) the 
\[
\frac{\hbar}{m_0}\gamma_{xz}=\langle\mathcal{Z}|p_x|\mathcal{X}\rangle
\]
term allowed by broken reflection symmetry in a heterobilayer, and (ii) the 
\[
\frac{\hbar}{m_0}\gamma_6=\left\langle\frac{\mathcal{X} + \mathrm{i}\mathcal{Y}}{\sqrt{2}} \bigg| {\bm{e}_-^* \cdot \hat{\bm{p}}} \bigg| \frac{\mathcal{X}' - \mathrm{i}\mathcal{Y}'}{\sqrt{2}}\right\rangle
\]
term (following the notations of Refs.~\cite{Kormanyos:2013dq,2053-1583-2-2-022001}) allowed by the threefold rotational symmetry of the monolayer; $m_0$ is the free electron mass. Microscopically, $\gamma_{xz} \ne 0$ can be related, for example, to the mixing of $\mathcal Z$-type $c+1$ conduction band $v$ with the valence band with invariant Bloch amplitude $\mathcal S$ of the same MoX$_2$ layer caused by local electric fields in the presence of the adjacent WY$_2$ layer in the heterobilayer, cf. Refs.~\cite{ochoaSpinorbitmediatedSpinRelaxation2013,PhysRevX.4.011034,mittenzweyManybodyRashbaSpinorbit2026} where Rashba effect has been discussed for monolayer semiconductors. Note that a non-zero $p_{\downarrow\uparrow}$ is possible in heterobilayers. In monolayers (in the absence of perturbations that break $z\to -z$ reflection symmetry), both $\gamma_{xz}\equiv $ and $\alpha'\equiv 0$. Hence, in monolayers, in agreement with symmetry considerations as the electric-dipole transitions between the spin subbands, are not allowed.

The transition rate under circularly polarized irradiation follows from Fermi's golden rule:
\begin{equation}
    \label{FGR:gen}
    W_{\downarrow\uparrow}^{\rm EDSR} = \frac{2\pi}{\hbar}\left(\frac{e|E_0|}{m_0\omega}\right)^2\sum_{\bm{k}}|p_{\downarrow\uparrow}|^2\delta(\hbar\omega + \epsilon_{\bm{k}\uparrow} - \epsilon_{\bm{k}\downarrow})f_{\bm{k}},
\end{equation}
where $E_0$ is the complex amplitude of the incident electromagnetic field, $\bm k$ is the electron in-plane wavevector (reckoned from the $K^+$ point of the Brillouin zone, $\epsilon_{\bm{k}\uparrow/\downarrow}$ are electron dispersions in the spin-orbit-split subbands, and $f_{\bm k}$ is the electron density. We assume that only the bottom conduction spin-subband is filled. We approximate electron dispersions in the effective mass model  
\begin{equation}
\label{parabolic:disper}
\epsilon_{\bm{k}\downarrow} = \frac{\hbar^2 k^2}{2m_\downarrow}  + \Delta_{SO}^c, \quad  \epsilon_{\bm k\uparrow} = \frac{\hbar^2 k^2}{2m_\uparrow},
\end{equation}
with $m_{\downarrow/\uparrow}$ being the effective masses and $\Delta_{SO}^c$ being the spin-orbit splitting of the conduction band spin subbands. The transition rate is given by
\begin{equation}
    \label{FGR:1}
    W_{\downarrow\uparrow}^{\rm EDSR} = {\frac{2\pi}{\hbar}D|p_{\downarrow\uparrow}|^2\left(\frac{e|E_0|}{m_0\omega}\right)^2\mathcal F(\hbar\omega,\Delta_{SO}^c)}.
\end{equation}
Here $D=|m^*|/(2\pi \hbar^2)$ is the electron density of states per spin, with the mass 
\begin{equation}
\label{m*}
    m^*=m_\uparrow  m_\downarrow/(m_\uparrow-m_\downarrow),
\end{equation}
The function $\mathcal F$ equals to unity if $\hbar\omega$ is between $\Delta_{SO}^c$ and $\tilde \Delta_{SO}^c = \Delta_{SO}^c+ \hbar^2 k_F^2/2m^*$ where $k_F$ is the electron Fermi wavevector (we consider the case of zero temperature for simplicity):
\[
\mathcal F(\hbar\omega,\Delta_{SO}^c) = \begin{cases}
\theta(\hbar\omega - \Delta_{SO}^c)\theta(\tilde \Delta_{SO}^c - \hbar\omega), \quad m^*>0,\\
\theta(\hbar\omega - \tilde\Delta_{SO}^c)\theta( \Delta_{SO}^c  - \hbar\omega), \quad m^*<0,\\
\end{cases}
\]
and $\theta(x)$ is the Heaviside $\theta$-function. The sign of $m^*$ is controlled by the relation between the effective masses $m_{\uparrow}$ and $m_{\downarrow}$. The latter depends on the order of spin subbands in the conduction band and, hence, the efficiency of the $\bm k\cdot \bm p$-mixing of the conduction and valence band states. In Mo-based monolayers the bottom conduction band spin subband and top valence band spin subbands have the same spin components as a result of an interplay between the atomic spin-orbit interactions on metal and chalcogen atoms~\cite{PhysRevB.88.245436,PhysRevB.93.121107,2053-1583-2-2-022001}. As a result, the $\bm k\cdot \bm p$-mixing for the bottom, $\uparrow$, subband is stronger, and $m_\uparrow<m_\downarrow$ in Mo-based monolayers~\cite{2053-1583-2-2-022001} and $m^*$ is negative. Equation~\eqref{FGR:1} shows that the THz absorption spectrum is rectangular and ranges from $\tilde \Delta_{SO}^c$ to $\Delta_{SO}^c$ in our case. The width of absorption spectra is controlled by $m^*$ and $k_F$ providing access to $m^*$ provided that electron density and, hence, $k_F$ is known.

Interestingly, the transition rate strongly increases as $m_\downarrow$ approaches $m_\uparrow$ but the range of frequencies where the transition is allowed decreases: It is because for equal masses the dispersion curves of subbands are parallel and the resonance condition $\hbar\omega = \epsilon_{\bm k\downarrow} - \epsilon_{\bm k\uparrow}$ is fulfilled for all values of $\bm k$ in the range of validity of Eqs.~\eqref{parabolic:disper}. The product $D\mathcal F(\hbar\omega, \Delta_{SO}^c) \propto \delta(\hbar\omega - \Delta_{SO}^c)$ where $\delta(x)$ is the Dirac $\delta$-function, and the transition occurs at a single spin-resonance frequency $\omega = \Delta_{SO}^c/\hbar$.

For comparison, we also calculate the magnetic-dipole transition rate between subbands $W_{\downarrow\uparrow}^{\rm M}$ driven by the Zeeman coupling~\eqref{Zeeman}. Calculations show that the ratio of the electric-dipole to the magnetic-dipole transition rates 
\begin{equation}
    \label{ratio}
    \frac{W_{\downarrow\uparrow}^{\rm EDSR}}{W_{\downarrow\uparrow}^{\rm M}} = \left(\frac{2c|p_{\downarrow\uparrow}|}{\Delta_{SO}^c}\right)^2.
\end{equation}
Taking, for the sake of estimation, $p_{\downarrow\uparrow} =10^{-4}\ldots 10^{-3} m_0 (\gamma/\hbar)$, where $\gamma\approx 2$~eV\AA$=3\times 10^{7}$cm/s~is the parameter describing the interband velocity~\cite{2053-1583-2-2-022001} in TMDC monolayers and $\Delta_{SO}^c=20$~meV, we have $W_{\downarrow\uparrow}^{\rm EDSR}/W_{\downarrow\uparrow}^{\rm M} \sim 30\ldots 3000$, i.e., the electric-dipole transitions strongly dominate over the magnetic-dipole transitions. Reduction of $|p_{\downarrow\uparrow}|$, i.e., by reduction of $\alpha$, $\alpha'$ or $\gamma_{xz}$ which, in principle can be done by applying strain and variation of the chemical composition of the layers may somewhat reduce the role of the electric-dipole transitions. We note, however, that Eq.~\eqref{ratio} shows general possibility of dominant magnetic transitions for cases of ${\Delta_{SO}^c} < 2c|p_{\downarrow\uparrow}|$.

The transition rate~\eqref{FGR:1} is calculated for the free electrons and can be observed in $n$-doped heterobilayers. If the intersubband transitions are studied in undoped case under the conditions of exciton photogeneration by additional optical illumination, Eq.~\eqref{FGR:1} can be used with natural replacements: the electron density $N_e$ should be replaced by the exciton density $N_{ex}$, the density of states $D$ should contain, instead of $m^*$, analogous combination of exciton translational masses, and $\Delta_{SO}^c$ should be replaced by the exciton `singlet-triplet' splitting $\Delta$ that contains additional contributions due to binding energy differences, the electron-hole exchange interaction, and, generally, light-matter interaction~\cite{ren2023control}. Recent experiments~\cite{wietekTHZ} have demonstrated `singlet-triplet' interexcitonic transitions by THz absorption in heterobilayers. It opens up a playground for studying EDSR in van der Waals heterostructures optically.

Above we considered only single THz-photon processes where the rate of transitions is proportional to $|E_0|^2$, i.e., to the THz radiation intensity. At elevated intensities multiphoton processes can be of importance~\cite{ganichev_book,keldysh_ion}. The analysis of such effects is an interesting problem for the future.

\subsection{Rashba Hamiltonian in heterobilayers}

The presence of a non-zero momentum matrix element $p_{\downarrow\uparrow}$ between the electronic Bloch functions of the spin subbands at the $K$ point of the Brillouin zone leads to another important feature: the emergence of linear-in-wave-vector spin-dependent terms in the effective Hamiltonian for electrons in heterobilayers~\cite{dyakonov_book,doi:10.1143/JPSJ.37.1325,vasko,bychkov84}.

Let us construct the corresponding Hamiltonian -- an analogue of the Rashba Hamiltonian -- by combining the 
$\bm k\cdot \bm p$ perturbation theory and the method of invariants. We introduce the Pauli spin matrices 
 $\bm \sigma = (\sigma_x,\sigma_y,\sigma_z)$, acting in the basis of the spin states $\Gamma_6$ ($\uparrow$) and $\Gamma_4$ ($\downarrow$). In the $C_{3}$ group, the matrix $\sigma_z$ turns out to be invariant and enters the Hamiltonian with a factor $-{\Delta_{SO}^c}/{2}$. This contribution is responsible for the spin splitting of the subbands within a given valley. Moreover, a term $\propto k^2 \sigma_z$ is possible that is responsible for the difference of the effective masses in the spin subbands, $m_\uparrow\ne m_\downarrow$. In addition, the $C_3$ symmetry allows linear-in-$\bm k$ terms of the form $\alpha(\sigma_x k_y - \sigma_y k_x)$.
 
The coefficient in front of them can be easily established by calculating the electron velocity operator matrix element between the $\Gamma_4$ and $\Gamma_6$ states using the effective Hamiltonian and comparing this result with the calculation in the 
$\bm k\cdot \bm p$-method~\eqref{momentum}. As a result, for electrons in the $K^+$ valley, we obtain 
\begin{multline}
\label{H:R}
\mathcal H(\bm k) = - \frac{\Delta_{SO}^c}{2}\sigma_z+ \frac{\hbar^2 k^2}{2\bar m} \hat I  + \frac{\hbar^2k^2}{4m^*}\sigma_z \\
+ \frac{\hbar p_{\downarrow\uparrow}}{m_0} (\sigma_x k_y - \sigma_y k_x),
\end{multline}
where $\hat I$ is the $2\times 2$ identity matrix, ${\bar m}^{-1} =( m_\uparrow^{-1}+m_\downarrow^{-1})/2 $, and $m^*$ is given by Eq.~\eqref{m*}. The Hamiltonian in the $K^-$ valley can be obtained from Eq.~\eqref{H:R} by the time reversal. The Hamiltonian~\eqref{H:R} is an analogue of the Rashba Hamiltonian~\cite{rashbasheka, rashba64} for transition metal dichalcogenide heterobilayers.

It should be mentioned that the calculation of the intensity of intersubband transitions with spin flip, presented by expressions~\eqref{FGR:gen} and \eqref{FGR:1}, can be performed directly using the Hamiltonian~\eqref{H:R}, in which the replacement $\bm k\to \bm k -(e/\hbar c) \bm A$ should be made, where $\bm A$ is the vector potential of the electromagnetic field. Naturally, such a calculation leads to the same result, Eq.~\eqref{FGR:1}.

Noteworthy, according to the estimates presented above, the range of $\bm k$-linear terms in Eq.~\eqref{H:R} $\hbar p_{\downarrow\uparrow}/m_0 \sim 0.1 \ldots 1$~meV\AA~ reasonably agrees with the $\bm k$-linear terms in classical semiconductor heterostructures~\cite{ganichevInterplayRashbaDresselhaus2014}. Thus, the rates of THz inter-spin-subband absorption and intraband, Drude-like, absorption can be comparable~\cite{khudaiberdievPolarizationresolvedElectronSpin2025}.

\section{Conclusion}

In conclusion, we have developed a comprehensive theory of intersubband spin-flip transitions in transition metal dichalcogenide heterobilayers. Our symmetry analysis reveals a fundamental distinction between monolayers and heterobilayers: while inter-spin-subband transitions in monolayers are strictly forbidden in the electric-dipole approximation, they become allowed in heterobilayers due to the reduced $C_3$ point symmetry of the valley. This symmetry reduction eliminates the distinction between in-plane components of true vectors and pseudovectors, enabling electric dipole spin resonance.

We have derived the selection rules for these transitions for all six high-symmetry stacking configurations of transition metal dichalcogenide heterobilayers, demonstrating that the electric-dipole-active transitions between conduction band spin subbands are a universal feature. Using a microscopic model based on spin-orbit-induced band mixing, we identified the specific interband matrix elements responsible for the finite momentum matrix element between the spin-split subbands. This matrix element is shown to govern both the strength of the electric dipole spin resonance  and the appearance of linear-in-momentum spin-splitting terms in the effective Rashba-like Hamiltonian for the heterobilayer.

Our calculations demonstrate that the rate of electric-dipole spin-flip transitions exceeds that of magnetic-dipole transitions by several orders of magnitude, making electric dipole spin resonance the dominant mechanism for terahertz absorption in these systems.

\acknowledgments

The authors are grateful to E. Wietek, A. Chernikov, and A.M. Shentsev for valuable discussions. This work was supported by RSF Grant No. 23-12-00142-Continuation.

\end{document}